# Extension of ATLAST/LUVOIR's Capabilities to 5 μm, or Beyond

Michael Werner,[a*] Mark Swain,[a] Gautam Vasisht,[a] Xu Wang,[a] Steven Macenka,[a] Avi Mandell,[b] Shawn Domagal-Goldman,[b] Joel Green,[c] Chris Stark[c]

[a]Jet Propulsion Laboratory, California Institute of Technology, 4800 Oak Grove Drive, Pasadena, CA, USA 91109
[b]NASA Goddard Space Flight Center, 300 E. Street SW, Washington, DC, USA 20546
[c]Space Telescope Science Institute, 3700 San Martin Drive, Baltimore, MD, USA 21218

**Abstract**. ATLAST is a particular realization of LUVOIR, the Large Ultraviolet Optical Infrared telescope, a ~10 m diameter space telescope being defined for consideration in the 2020 Decadal Review of astronomy and astrophysics. ATLAST/LUVOIR is generally thought of as an ambient temperature (~300 K) system, and little consideration has been given to using it at infrared wavelengths longward of ~2 μm. We assess the scientific and technical benefits of operating such a telescope further into the infrared, with particular emphasis on the study of exoplanets, which is a major science theme for ATLAST/LUVOIR. For the study of exoplanet atmospheres, the capability to work at least out to 5.0 μm is highly desirable. Such an extension of the long wavelength limit of ATLAST would greatly increase its capabilities for studies of exoplanet atmospheres and provide powerful capabilities for the study of a wide range of astrophysical questions. We present a concept for a fiber-fed grating spectrometer which would enable R = 200 spectroscopy on ATLAST with minimal impact on the other focal planet instruments. We conclude that it is technically feasible and highly desirable scientifically to extend the wavelength range of ATLAST to at least 5 μm.

## 1 Introduction—ATLAST, LUVOIR, and the Need for Infrared Capabilities

Four distinct, if not totally independent, threads point towards a Large (~10 m class) UV-Optical-IR telescope in space (a.k.a. LUVOIR) as an important and exciting new astronomical facility for the not-too-distant future. First, the recently published 30-year road map for NASA Astrophysics, "Enduring Quests, Daring Visions", identified LUVOIR as a high priority future venture. Second, NASA's Astrophysics Division has designated LUVOIR as one of four mission concepts to be studied in depth as preparation for the 2020 Decadal Review of Astronomy and Astrophysics. Third, such an observatory, referred to as HDST for High Definition Space Telescope, is the centerpiece of an AURA-sponsored study on the Future of UVOIR Space Astronomy beyond JWST[1]. Finally, a JPL/GSFC/STScI team has studied such a system, ATLAST, which is described in companion papers within this volume[2]. ATLAST and HDST can be thought of as possible implementations of LUVOIR. LUVOIR itself will be defined by a NASA-selected Science and Technology Definition Team over the next three years.

Neither ATLAST nor HDST embraces the thermal infrared; both were envisioned as operating at room temperature. HDST baselines a long wavelength limit of 2 μm, with an "extended goal" in the 3–5 μm range. Similarly, ATLAST expresses its long wavelength limit as 1.8 μm, with a stretch goal of 2.5 μm, and an understanding that the telescope may be instrumented for measurements out to 5 μm (or perhaps beyond) but that this longer wavelength application will not levy any requirements upon the telescope.

In this paper, we present the scientific rationale and technical implications for the use of LUVOIR to at least 5 μm, arguing that this wavelength range should become the baseline for LUVOIR. To limit the scope of this investigation, we consider primarily the benefits of such an extension to the study of exoplanets. We do this because both ATLAST and HDST have made the study of exoplanets a major pillar of their scientific case. In the following discussion we refer to the telescope as ATLAST, because the work described here was supported by the ATLAST study, but going forward the arguments detailed below should be brought to bear on the definition of LUVOIR. Many of the arguments introduced here could also be applied to the study of the Habitable Planet Explorer (HabEx), which is a second mission being studied by NASA in preparation for the 2020 Decadal Review. We discuss both the baseline ATLAST (9.2 m, T = 293 K) and its enhanced cousin (12 m, T = 273 K).

The rationale for operating ATLAST out to at least 5 μm may be summarized as follows. One can be confident that, whenever ATLAST launches, the study of exoplanet atmospheres will be an exciting and crucial scientific discipline. Numerous molecular species likely to be important probes of exoplanet atmospheres have bands in the infrared, including $H_2O$, $CO$, $CO_2$, $CH_4$, $NH_3$, and $O_3$. Many of these same species are abundant in the Earth's atmosphere and difficult to access from the ground, but are easily studied by a space telescope such as ATLAST. In addition, recent developments suggest that atmospheric hazes or clouds may be common in exoplanet atmospheres and that they can interfere with our ability to determine molecular abundances using optical and near infrared measurements only[3]. Observations further into the infrared have the potential to improve constraints on molecular abundances by either punching through the haze or better defining its properties and extent. As a general principle, the study of exoplanet atmospheres will benefit greatly from inclusion of the widest range of molecules over the widest wavelength range. We propose the extension of ATLAST's capabilities in this spirit. ATLAST's infrared capabilities will inform our overall understanding of exoplanet atmospheres and, in ways which may be difficult to foresee at present, support our quest to identify habitable or potentially habitable planets through atmospheric studies.

Combined light spectroscopy and photometry of the atmospheres and surfaces of transiting planets in both transit (when they pass in front of the star) and eclipse (when they go behind), pioneered by Spitzer and Hubble and to be carried forward by JWST, will continue to be a powerful tool for exoplanet studies. WFIRST, equipped with a coronagraph, will pioneer extensive direct imaging and spectroscopy from space and might also be used for combined light transit spectroscopy. ATLAST will advance this science even further, observing targets discovered by both ground- and space-based surveys, in direct and, as proposed here, combined light spectroscopic modes. Of course, an infrared-capable ATLAST could also carry out exciting studies of targets other than exoplanets.

## 2   Use of a Warm Telescope for Infrared Astronomy from Space

The primary benefit of a cryogenic space telescope for infrared studies—high sensitivity—is well known. However, sensitivity is not the only rationale for an infrared space telescope. Here we consider instead the benefits of an "ambient temperature" (say 273–300 K) space telescope, in comparison to a comparable- or even larger-sized telescope at a mountaintop observatory. For present purposes, these benefits come in four categories:

a. Access to the entire infrared spectrum. As mentioned above, the very molecules of greatest interest for the fundamental study of exoplanet atmospheres, are present in the Earth's atmosphere as well, often cloaking the very features we wish to study. From outside the Earth's atmosphere, we can access the entire infrared spectrum and all infrared-active molecular species. This may include important species with critical diagnostic features longward of 5 μm, as well as the prominent species with spectral features in the 2–5 μm band discussed here.

b. Stability. The study of exoplanet spectra using combined light spectroscopy requires achieving a measurement precision measured in parts per million for the most difficult cases. Although achieving such precision on a large, warm telescope in space might not be trivial (20 ppm has been demonstrated with Hubble WFC3 IR grism measurements), it would certainly be easier than on a similar telescope on the ground, looking through the warm and turbulent atmosphere.

- c. Sensitivity. The large space telescope should achieve higher sensitivity than a comparably sized and similar temperature telescope on the ground because of the absence of atmospheric absorption and emission and the routine achievement of diffraction-limited performance.
- d. Clear Skies and Long Observations. Many exoplanet observations—for example, transits, eclipses, and phase curves—will be of long duration and have timing constraints set by the orbital period of the planet so that a particular phenomenon might be observable only a few times per year. A space telescope, operating always under cloudless skies and easily able to observe a particular target for days rather than hours, provides great practical advantages for this type of study.

The scientific, technical, and functional considerations outlined above motivated us to explore extending the wavelength range of ATLAST to 5 μm, without imposing a constraint on the telescope temperature, but by adding a suitable scientific instrument. We begin by exploring the scientific benefits of this extended wavelength coverage. In Appendix A, we present a strawman design for an infrared spectrometer, covering the 1–5 μm wavelength range with resolving power R = 200, which could execute the science described below. Five microns is not an arbitrary cutoff, because it reflects current detector technology; five micron cutoff HgCdTe has been extensively developed for use on JWST. Nevertheless, a more thorough study should explore longer wavelengths, noting, for example, that the 5–8 μm spectral interval is largely inaccessible from the ground and that McMurtry et al.[4] have demonstrated excellent performance in HgCdTe with a cutoff wavelength around 10 μm at temperatures 35-to-40K achievable through passive cooling.

# 3 Exoplanet Science—Atmospheric Characterization with Transit Spectroscopy

## 3.1 Telescope Size Can Trump Background Noise

Transit spectroscopy and photometry study the atmosphere of a transiting exoplanet by measuring the spectral features which the atmosphere imprints on the starlight which filters through it en route to us (the scientific literature contains numerous recent examples of both transit and eclipse spectroscopy and photometry[3,5–7]). Transit spectroscopy requires observations of the brightest stars, for which the main noise source, even with a warm telescope, is the photon noise of the star itself. For these observations of bright stars there is nothing gained, in principle, by cooling the telescope (see Fig. 1 below). In Fig. 1, we consider the time taken (or inverse sensitivity) for three telescope configurations to detect a similar small (10 ppm at R = 200) and narrow feature at 4 μm, as a function of the stellar magnitude at that wavelength. The configurations are baseline ATLAST (9.2 m, T = 293 K), the larger and somewhat cooler "stretch goal" ATLAST (12 m, T = 273 K), and JWST (6.5 m, cryogenic telescope). As suggested above, ATLAST is more sensitive than JWST for bright host stars, with the 12 m ATLAST telescope taking ~4× less time to make the same observation for stars brighter than [L]~9, as is to be expected in the stellar photon limited case. For transit spectroscopy, a large warm telescope becomes more powerful than a smaller but cooler one. This is exactly the reverse of the situation which applies for long-wavelength, low-to-moderate spectral resolution measurements of the emission from individual stellar sources, where small cryogenically cooled telescopes like IRAS, ISO, Spitzer, and WISE, outperform even 10 m-class ambient temperature telescopes by huge factors beyond ~3 μm.

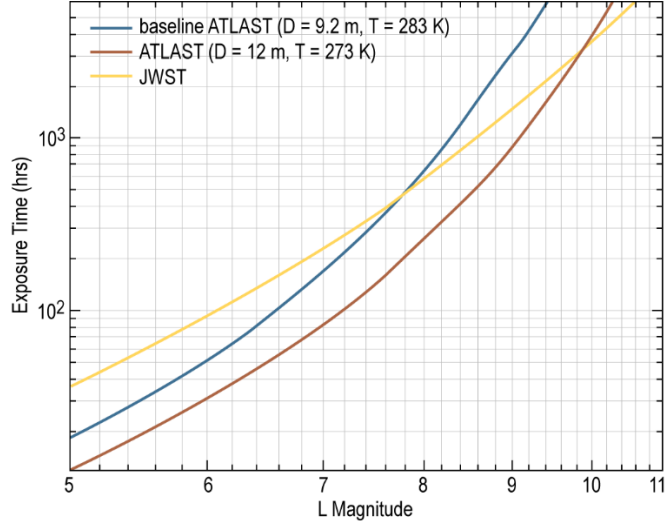

**Fig. 1** The time taken (or inverse sensitivity) for three telescope configurations to detect with 5:1 signal-to-noise a small (10 ppm at R = 200) and narrow feature at 4 µm due to a transiting exoplanet, as a function of the stellar magnitude at that wavelength. The configurations are baseline ATLAST, the larger and somewhat cooler "stretch goal" ATLAST, and JWST. ATLAST is more sensitive than JWST for bright host stars, with the 12 m ATLAST telescope taking ~4× less time to make the same observation, as is to be expected in the stellar photon limited case. For the ATLAST configurations, a total photon throughput of 0.35, a (single mode) background with emissivity of 0.2 is assumed; these parameters are commensurate with those of the instrument described in Appendix A. For JWST, the actual expected performance of the NIRCAM grism[8] is adopted.

A transiting planet in a circular orbit undergoes an eclipse as it goes behind the star half an orbit later. If the planet produces appreciable thermal emission or even scattered light, the total radiation received from the star + planet system drops during this eclipse; therefore the planet-related flux can be determined by comparing in-eclipse versus out-of-eclipse measurements. The depth of the eclipse signal will be less than that of the transit; it depends on the temperature ratio of the star and the planet as well as on their relative areas. However, an atmospheric absorption feature might be more readily detected because the entire disk of the exoplanet, and not merely the relatively thin atmosphere, contributes to its observability, although the temperature structure of the exoplanet's atmosphere has a thumb on the observability scale as well. Because of these complexities, eclipses will not be considered in any detail in this paper, but ought to be included in any further analysis of the science return from adding infrared capability to ATLAST. The same argument which shows that a large warm telescope outperforms a smaller but colder one for transit measurements of bright stars shows that the former wins out for eclipse measurements over the same magnitude ranges as well. Together, the study of exoplanet transits and eclipses, together with the intervening phase curve, are referred to as combined light spectroscopy or photometry.

*3.2 The Current State of Play in Combined Light Studies of Exoplanets*

Exoplanet direct imaging is a major driver of the ATLAST concept, which includes coronagraphy and/or the use of an external star shade to obtain direct images and low-resolution spectra of exoplanets. The infrared functionality discussed here provides a powerful complement to this direct imaging work through the use of combined light studies. Observations of transiting exoplanets with ATLAST can yield an enormous scientific return due to the combination of a large aperture and an abundance of targets.

Extraordinary space-based measurements transit and eclipse measurements from Hubble, Spitzer, and Kepler have dominated the emerging field of exoplanet atmosphere characterization by combined light spectroscopic and photometric measurements. We have learned much from this large and rapidly growing body of work. Some of the highlights[5–7,9–11] include (T = result from transit study; E from eclipse; P from phase curves, which follow the planet from transit through eclipse and back):

- Water is common and is present in a majority of Hubble transmission spectra [T]
- Heat transport from dayside to nightside varies enormously in efficiency [P, Spitzer and Hubble]

- Large-scale zonal winds are present and can, in some regions of the atmosphere, dominate the local energy balance [P, Spitzer]
- Clouds and aerosol hazes are frequently present to varying degrees [T, Hubble and Spitzer]
- Some planets have bow shocks produced by significant magnetic fields [T, Hubble]

While these and other aspects of exoplanet atmospheres give significant insight into the conditions and processes of exoplanet atmospheres, we still do not have a good understanding of the controlling variables. For example the bulk composition C/O ratio could play an important role in determining the strength of the water feature in an exoplanet atmosphere; however, in spite of significant effort, basic questions such as the typical C/O ratio for hot-Jupiter exoplanets are still not understood. Improving the observation quality and quantity is essential. Fortunately, the community can look forward to the arrival of JWST in 2018, with significantly enhanced capability, in terms of both wavelength coverage and sensitivity, over what is available today[12]. Just as HST and Spitzer are setting the stage for exoplanet studies from JWST, JWST and WFIRST will further advance the field, defining science themes and identifying targets of interest for study from ATLAST. In addition there have been white papers and active proposals to both NASA and ESA to fly a modest-sized dedicated mission to conduct an exoplanet transit characterization survey.

*3.3 There will be Targets*

We expect that numerous stars as bright as those shown in Fig. 1 which are home to transiting planets will have been identified by the time ATLAST is launched more than a decade from now. The K2 mission is operational and finding bright transiting exoplanet systems today. The TESS mission, launching in 2017, is expected to find thousands of transiting planets around host stars ranging in I-band brightness from 12th magnitude to brighter than 5th magnitude. TESS will be followed by the ESA M3 PLATO mission, launching in 2024, which will find thousands of exoplanets around bright stars, as discussed further below. In the post-JWST era, ATLAST can provide a critical capability to follow up these discoveries with detailed characterization. Throughout this period, ground-based transit and radio velocity surveys such as Mearth, WaSP, and HARPS, will also identify planets around bright stars suitable for study from ATLAST.

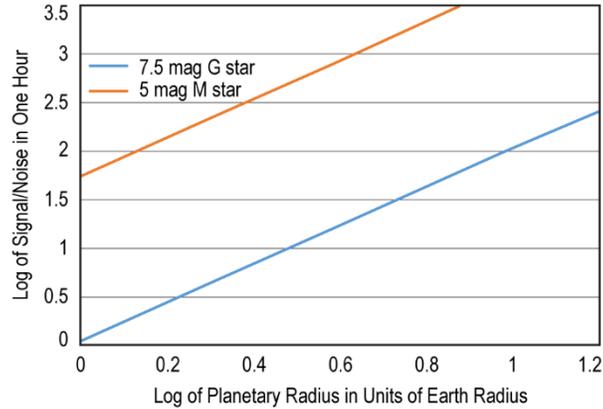

**Fig. 2** Signal/noise for each resolution element on the transit in one hour for planetary transits at 4 μm. A 9.2 m 293 K telescope and the R = 200, fiber-fed spectrograph described in Appendix A are used for the measurement.

In particular, the PLATO mission, to be launched by ESA in 2024 for a six+ year mission, will find scores of planets for study by ATLAST. Based on information provided by David Brown of the PLATO team, Plato should find over 3400 transiting planets around stars brighter than 11th magnitude in the visible. From this, we estimate that PLATO could find transiting planets around ~130 K-type main sequence stars (and dozens of G and M stars) brighter than 7.5 mag at 4 μm; many of these planets should lie in or near the habitable zone. The predicted S/N per hour vs. planet size for the some of the most readily studied exoplanets which PLATO might discover is shown in Fig. 2. We take S/N > 100 as target performance, because only a portion of the total transit signal is contributed by the atmosphere. This is readily achieved in a single transit for Jovian planets with R ~ 10 $R_{Earth}$. A handful of smaller planets will orbit stars of ~5th mag, so high S/N is achieved in one transit even for planets smaller than 2 Earth radii. (Stars of this brightness can be observed with ATLAST and the R = 200 spectrometer described in the Appendix without saturating the detectors, provided that frame rates faster than 1 Hz are achievable.) Even for fainter stars, degrading resolution and coadding transits should allow useful S/N to be achieved in a reasonable time to below 2 Earth radii. Note that the improved sensitivity for the transit measurement of the planet orbiting the M star versus that orbiting the G star (Fig. 2) results (in roughly equal amounts) from both the higher apparent brightness of the M star and its smaller area.

*3.4 Abundance Constraints—An Example*

A critical performance metric for characterizing exoplanet atmospheres via spectroscopy is the degree of constraint the measurements provide on the abundance, also termed the atmospheric mixing ratio, of the species detected. We use this as a first-order estimate of detectability which does not yet account for differences in concentration versus altitude. That said, it is useful for providing comparisons between similar observational scenarios and analysis approaches as we do below.

To evaluate the effects of the ATLAST wavelength range on the constraints that transit spectroscopy with ATLAST might provide, we have simulated transmission spectra (we use transmission spectrum and transit spectrum interchangeably in what follows) based on an atmospheric forward model, added noise based on an instrument model, and then performed a multiparameter Markov Chain Monte Carlo (MCMC) retrieval using the CHIMERA code. (We thank a referee for pointing out to us a recent paper by Greene et al.[13] which, for the specific case of JWST, shows the importance of observations longward of 2.5 µm for characterizing exoplanet atmospheres). For small planets, the viability of transmission spectroscopy depends critically on the planet size, the parent star brightness, and the scale height of the planetary atmosphere. As an illustrative example, we considered the case of a 2.6 $R_{Earth}$, 6.1 $M_{Earth}$ transitional Super Earth/sub-Neptune planet with a 600 K atmosphere and assumed solar metallicity, thermal equilibrium chemistry, and a clear atmosphere, orbiting a K star. This atmosphere, which we refer to below as the low molecular weight case, is dominated by $H_2$ and has a mean molecular weight of 2.3. As a comparison, we included as well a high molecular weight case, which we generated by stripping all the $H_2$ out of the low molecular weight case, but not redoing the atmospheric chemistry. The resulting atmosphere, with a mean molecular weight of 28, may be somewhat unrealistic chemically, but it serves as a useful test case for present purposes; of course, with the higher mean molecular weight it will have a lower scale height, and a smaller atmospheric signal in transit, than does the low mean molecular weight planet.

Based on extensive numerical simulations, we have determined the stellar H band magnitude at which the abundance of water and methane can be determined in 24 hours of observation with a total ± 1 sigma uncertainty of less than one order of magnitude; one example is shown and described in Fig. 3. The results of these experiments are presented in Table 1, which shows that with 0.4–5.0 μm spectral coverage, ATLAST has the potential to determine the abundance of water and methane in many TESS/K2/PLATO targets. If the spectral coverage were limited to 2.5 μm, brighter stars are required, so substantially fewer targets would become available for study. This demonstrates the benefits of the enhanced wavelength range which we are proposing. The table shows that the benefits of the extended wavelength coverage are particular marked for the high mean molecular weight atmosphere, which corresponds approximately to an Earth-like atmosphere. In this case, a very bright star ([H] < 6.3) is required for abundance determinations with less than an order of magnitude uncertainty; stars this bright which harbor transiting planets may be rare statistically. As expected, the lower mean molecular weight atmosphere, with its larger scale height, would be much more readily observed. Several recent papers[14,15] have highlighted the effects of refraction in the lower part of an exoplanet's atmosphere on our ability to study abundances using transit measurements. The importance of this effect depends on both the size of the star and the star-planet distance, and it will have to be taken into consideration in more sophisticated predictions of the type presented below.

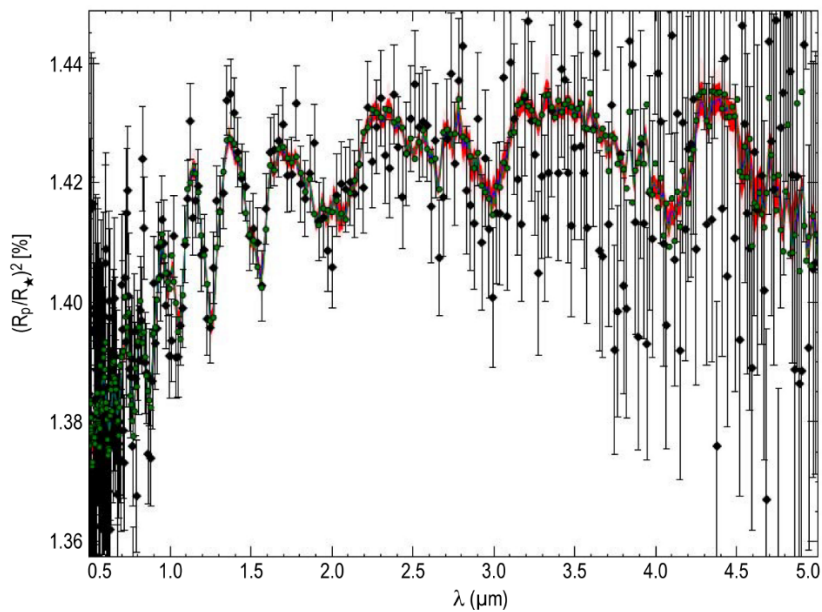

**Fig. 3** An example of abundance estimation from transit spectroscopy. The figure shows a simulated transmission spectrum of a Super Earth/sub-Neptune with high mean molecular weight (mmw = 28) atmosphere orbiting an $H_{mag}$ = 8.3 K dwarf host star. Six transits are coadded—leading to a total integration time of 24 hours—to produce this spectrum with a 9.2 m ATLAST and nominal instrumental parameters. The forward model and retrieval used in this analysis are part of the CHIMERA package developed by Mike Line[16]. The plotted spectrum was obtained with a resolving power of 1000; if the resolution were degraded to the R = 200 level obtainable with the spectrometer described elsewhere in this paper, the abundance determinations for the broad molecular bands would not be materially affected. As summarized in Table 1, spectra of this quality extended to 5 μm and analyzed with CHIMERA can constrain the abundances of molecules like $CH_4$ and $H_2O$ in exoplanet atmospheres with a total uncertainty less than an order of magnitude.

Table 1 Limiting Magnitude for Determination of Molecular Abundances with an Overall Uncertainty of Less Than One Order of Magnitude for a 2.6 $R_{Earth}$ Super Earth/sub-Neptune (T ~ 600 K) orbiting a K star.

|  | Molecule | 0.4–2.5 μm | 0.4–5.0 μm |
|---|---|---|---|
| Mean mol wt 2.3 | $H_2O$ | $H_{mag} < 10.3$ | $H_{mag} < 11.3$ |
|  | $CH_4$ | $H_{mag} < 10.3$ | $H_{mag} < 12.3$ |
| Mean mol wt 28 | $H_2O$ | $H_{mag} < 6.3$ | $H_{mag} < 8.3$ |
|  | $CH_4$ | $H_{mag} < 6.3$ | $H_{mag} < 8.3$ |

## 4 Studying a True Earth Analog

### 4.1 Habitability and Biosignatures

A major thrust of exoplanet studies is, of course, the push to identify and study potentially habitable Earth-like planets. As indicated in a companion paper by Domagal-Goldman et al.[17], recent work has shown that simply identifying $O_2$ and $O_3$ in a particular atmosphere is not adequate. Other molecular species, notably $CH_4$ but also $N_2$, CO, and $CO_2$, may have to be studied in order to rule out or confirm the potential habitability of a particular planet which shows $O_2$ and $O_3$. Understanding the prevalence of these species (all of which have strong spectral features longward of 2.5 μm) in exoplanet atmosphere, thus takes on special interest, even if may not in general be possible to search for all of these molecules in one and the same planetary atmosphere.

### 4.2 Transit Spectroscopy

The detection of biomarkers in the atmosphere of a true Earth analog presents major challenges. This is a difficult detection, so we present it as a limiting case. The average methane volume mixing ratio in the Earth's atmosphere is quite low ($1.7 \times 10^{-6}$ mol/mol), a factor 200 less than $CO_2$, but it is the major source of departure from thermodynamic equilibrium if the atmosphere is considered as a closed system (without oceans and rocks). The strong methane 3.4 μm feature will be blended with a water absorption at low R, but nonetheless contributes up to 10 km (a little more than the 8 km scale height) of extra absorption in the transmission spectroscopy of an Earth. The weaker methane features at shorter wavelengths are yet more heavily blended. In atmospheres with high cloud coverage, the strong IR molecular features absorb at altitudes above the cloud deck, which is a major advantage of long wavelength spectroscopy.

We have computed the transit spectrum of an Earth analog orbiting a late type star, exploiting the well-known advantages conferred by M stars when considering transit measurements; in addition, the refraction effect mentioned above is much less of a concern for transit spectroscopy of HZ Earths around M stars than around solar type stars[14]. Space missions such as K2, TESS, and PLATO, as well as ground-based surveys targeting M stars, will scour nearby bright stars for transiting terrestrial planets orbiting M dwarf hosts. Assuming that the occurrence of habitable zone rocky planets is 0.2 around late type stars, the nearest M-dwarf with a favorable transit (as opposed to a short glancing transit) will be about ~15 pc away; considering the space density of more massive stars and similar occurrence rates, host K and G main sequence dwarfs will be at distances of ~30 and 50 pc, respectively.

In Fig. 4 we show a simulated ATLAST R = 20 transit spectrum of an Earth transiting an M4V star (distance 15 pc, $L_{mag}$ = 7.7) in the 2–5 μm region; the underlying model simply uses an Earth transit spectrum and the instrument prescription described in this paper, for which the spectral resolution can be degraded from 200 to 20 with a sqrt(10) increase in S/N per resolution element. While the total time required for the spectrum shown in Fig. 4, ~60 transits or 120 hrs, is large, it is not outrageously so (two exoplanets have had similar amounts of Hubble time awarded). The orbital period of a habitable zone planet around our M4 star is 16 days, so about 2.5 years would be required to obtain spectra similar to that shown in Fig. 4.

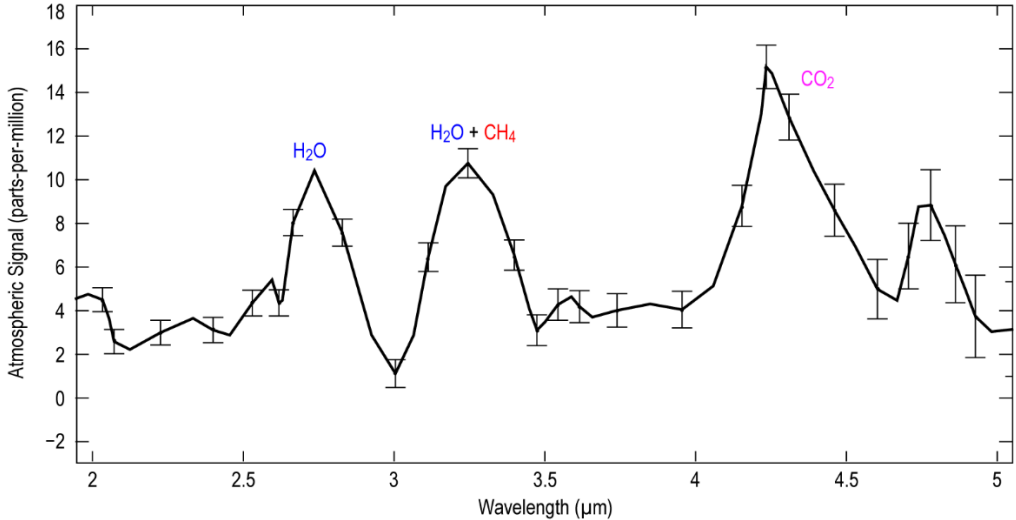

**Fig. 4** A 2–5 μm R ~ 20 model transmission spectrum of a 1 $R_{Earth}$ habitable zone planet with an Earth-like oxidizing atmosphere transiting an M4V star ($R_{star}$ = 0.32 $R_s$, $T_{star}$ = 3100 K) which is 15 pc from Earth. The 1-sigma errors per wavelength bin, obtained with the baseline ATLAST design (D = 9.2 m, T = 273 K), are plotted along with the model. The total integration time is 120 hours in transit, corresponding to 60 transits of about 2 hours each. Features of $H_2O$, $CH_4$, and $CO_2$ are detected. $CH_4$ is blended with $H_2O$ at ~3.3 μm and determining its independent presence and total column would require full atmospheric retrieval; in the Earth's atmosphere, $CH_4$ and $H_2O$ contribute roughly equally to the 3.3 μm feature. A total throughput of 0.35 and an emissivity of 0.2 is assumed here. The y-axis is labelled in parts per million relative to the total signal produced by the star. The overall depth of the transit for this case would be about 800 ppm; the features shown in the spectrum would be manifest as changes in the transit depth with wavelength.

To show how ATLAST will be able to distinguish between exoplanets of varying composition, we present in Fig. 5 (to be compared with Fig. 4) the transmission spectrum of a sub-Neptune orbiting the same M4V star.

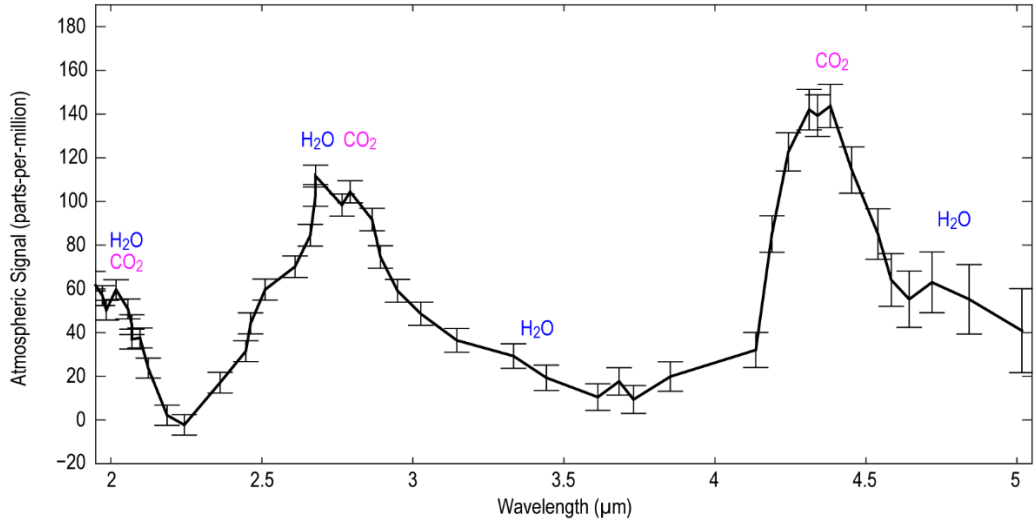

Fig. 5 A hypothetical water (80%) and $CO_2$ (20%) (mean molecular weight of 23) sub-Neptune planet transiting the same M4V star (15 pc). This planet has a radius of 2.7 $R_{Earth}$ and a mass of 6.5 $M_{Earth}$. The atmospheric temperature is ~500 K, so it is a bit closer to the star than is the Earth analogue shown in Fig. 5. Error bars indicate the 1-sigma precision that could be obtained along the spectrum per R = 100 spectral bin in 6 hours, which would require observations of a handful of transits.

## *4.3 Direct Imaging Spectroscopy*

To this point, we have considered only transit spectroscopy as a means of characterizing exoplanet atmospheres. However, the impetus to study true Earth analogs is sufficiently strong that it is useful to consider how such characterization could be carried out with direct imaging, either with a coronagraph or with a starshade starlight suppression system, on an ambient temperature ATLAST.

The strategy to observe biosignatures in the atmospheres of exoEarths includes near-IR capabilities at low resolution out to 3.4 μm. The case for detection includes a comparison of molecular oxygen, ozone, carbon monoxide, and carbon dioxide tracers at wavelengths shorter than 2.4 μm (recall, however, that CO and $CO_2$ have their strongest bands between 4 and 5 μm).

However, without methane at 2.3 or 3.4 μm, it will be difficult to tell the difference between biotic and abiotic processes[17].

Constraining the biosignatures to biotic processes for even a single system would add great value to the experiment conducted by ATLAST. The small additional cost of an infrared instrument such as that described below is an excellent tradeoff against the cost of a full followup mission, in addition, it could guide the development of such a mission.

For this exercise, we assumed a diffraction-limited 12 m diameter unobstructed primary, a coronagraph with 2 λ/D inner working angle and $10^{-10}$ contrast sensitivity, noiseless detectors, and total system throughput of 20%. We calculated the numbers of Habitable Zones (HZs) around nearby stars that are observable at 2.3 and 3.4 μm, and for which a single R = 70, SNR = 5 per spectral channel spectrum of a true Earth analog could be obtained in less than 100 days of exposure time; higher SNR is highly desirable for quantitative assessment of biosignatures. We assumed that the planet is an Earth-twin initially discovered at shorter wavelengths, and that spectra are obtained when the planet is at quadrature. We ignored the thermal emission of the planet and emission from an exozodiacal dust cloud (exozodi), but included thermal emission from the telescope's primary and secondary mirrors (adopting an emissivity of 4% for each mirror) in our exposure time calculations.

Figure 6 shows the cumulative number of accessible HZs at 2.3 μm (orange) and 3.4 μm (red). This figure illustrates the number of HZs that have individual spectral characterization times less than the adopted 100-day limit—not all can actually be observed within the 100-day limit; and, of course, not all of them would necessarily be home to terrestrial planets. A 300 K telescope prohibits the characterization of any planets, even with 100 days of exposure time, due to the telescope's blackbody emission. Moderate cooling of the telescope to ~275 K permits the observation of the Alpha Centauri binary system. This nearby system can be explored almost completely. To access more than a few HZs, one must cool the telescope—as well as the coronagraph if one were used for starlight suppression—to ~200 K.

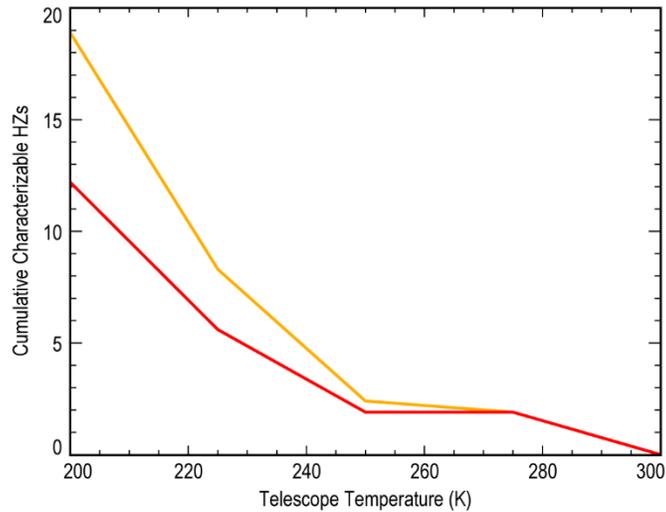

**Fig. 6** Number of HZs potentially containing Earth analogs which could be characterized at 2.3 μm (orange) and 3.4 μm (red) as a function of telescope temperature, assuming an upper limit of 100 days of exposure time on each system. Cooling to ~275 K provides access to the Alpha Cen binary system. Cooler temperatures are required to reduce the exposure times of other systems. A 12-m aperture telescope is assumed.

*4.4  Cool is Cool*

The foregoing shows that, for an ambient temperature telescope even as large as 12 m, searching for biosignatures at wavelengths beyond 2 μm will be more plausible using transit spectroscopy than using direct imaging. In no case, however, is measuring such biosignatures in a true Earth analog a simple thing in this wavelength range. Although a warm ATLAST will not yield a large bounty of exoEarths characterizable in the IR by direct imaging, the chance to confirm the few top priority systems may be worth the small additional cost. However, our estimates (see also Fig. 6) suggest that it does require cooling to close to 200 K to access the critical spectral lines longward of ~2 μm.

Two other possibilities exist. One is that one and the same habitable zone planet will be directly observable by both the coronagraph and the combined light spectrograph on ATLAST. It is unlikely that a randomly chosen exoplanet seen in direct imaging will also transit the star. However, if enough such planets meet our criteria of rocky and in the habitable zone, then a handful of transiting examples may be available to us. Another possibility is that the transit spectroscopy technique can be extended into the visible, so a complete ~0.5-to-5 μm spectrum of a single transiting planet can be obtained which covers all important biomarkers.

## 5  Beyond Exoplanets

This study foresees the operation of a diffraction limited ~10 m diameter telescope in space equipped with a modest resolution spectrograph covering wavelengths out to at least 5 μm. The scientific applications of such a system go far beyond the studies of habitability and exoplanet atmospheres discussed above. As one of many possible examples, we can also consider its utility for probing the formation of planetary systems.

The spacecraft stability will allow for exquisite time series analyses of IR spectral features. Although we currently think of solid-state features in mid-IR disk spectra as static snapshots of dust and ice processes, growth, destruction, and chemical modification occur on single mission timescales, particularly in the more active inner disk regions of accreting young stars. In particular, eruptive accretion events are uncommon but excellent probes of disk chemistry, and the opportunity to catch one during the lifetime of ATLAST will enable studies of inner disks. Although JWST will cover this area very well, the increased spatial resolution of ATLAST will provide insights into radial transport, especially in light of high-resolution ALMA observations of systems like HL Tau (Fig. 7).

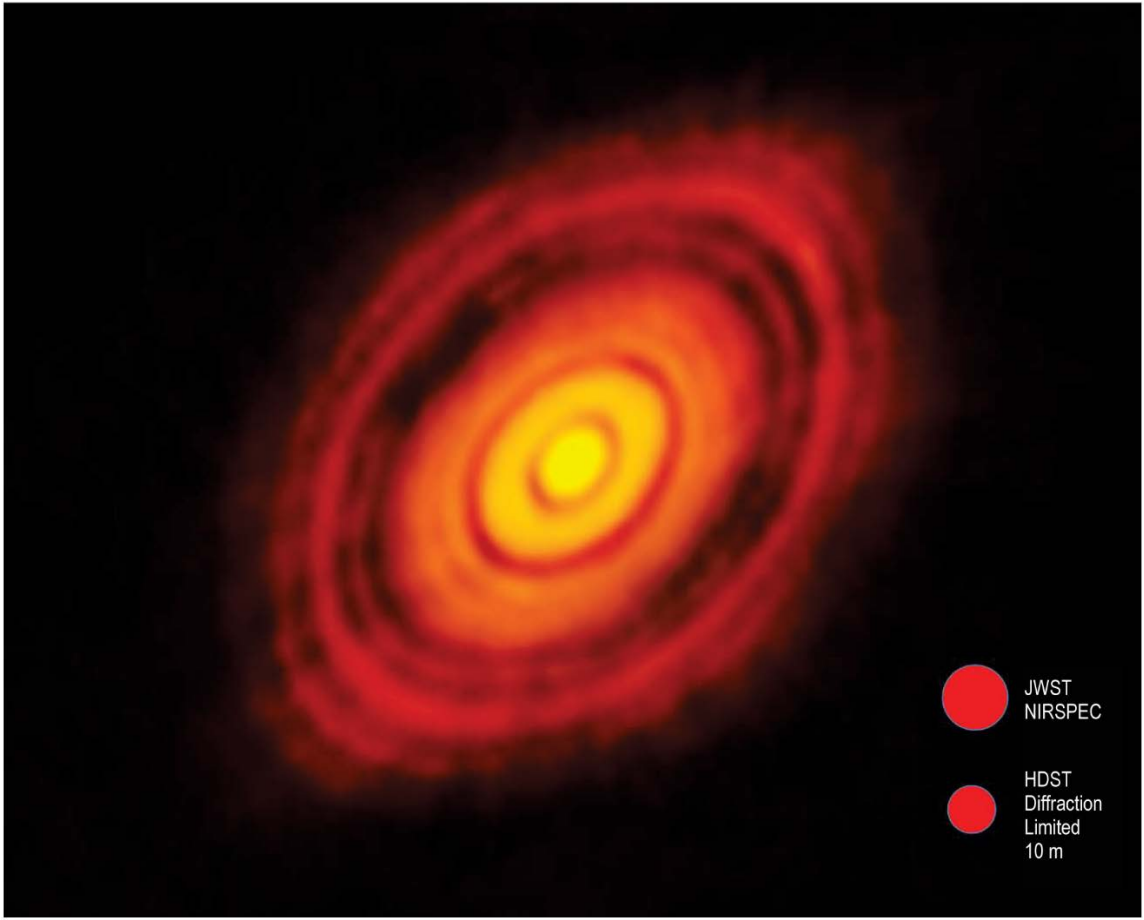

**Fig. 7** HL Tau as seen by ALMA (https://public.nrao.edu/news/pressreleases/planet-formation-alma), compared with approximate JWST and HDST beamsizes at 3 μm.

Additionally, circumstellar ices such as $H_2O$ (3.0 μm) will appear in scattered light[18]. Although $H_2O$ ice can be observed from the ground, $CO_2$ ice (4.3 μm) cannot, and a low resolution spectrograph is the perfect instrument to observe $CO_2$ ice. For example, an increase in spatial resolution by a factor of 1.5 over JWST (assuming a 10 m aperture) decreases the resolved scale in typical low mass star forming regions like Ophiuchus at 120 pc or Taurus at 140 pc (regions containing numerous proto-solar analogues) from ~8–16 AU down to 6–12 AU. This shifts the innermost pixel fully within some chemical ice/snow lines for some systems. The LUVOIR beam would fall well within the innermost detected gap in the HL Tau disk. Furthermore, 3–5 µm is the optimal range to track planets embedded in disks. The ability to observe a ground-inaccessible snowline in $CO_2$ scattered light at unprecedented resolution would inform numerous models of planet formation. Note also that solids (ices, dust) redistribute on time scales of hundreds to thousands of years, following snow-lines such as those of CO[19].

Thus a key motivation for the infrared extension of ATLAST would be observation of changes in chemical composition as conditions change real-time in a disk, either with solid-state features (e.g., $H_2O$ or $CO_2$ ice, or the PAH band at 3.3 µm) or with unresolved spectral lines tracing accretion shocks (e.g., [Fe II] @ 1.644 µm) or outflow-driven shocks ($H_2$, CO) within the disk and the surrounding envelope. The observations would also probe the launching point of outflows and jets. They would provide > 2× better areal resolution than JWST and complete coverage of the spectral regions inaccessible to a ground-based ELT. Further information about the evolution of these systems on dynamical time scales will come from comparing ATLAST, ALMA, and JWST observations of the same protoplanetary systems, separated in time by perhaps a decade.

Of course, protoplanetary disks systems such as HL Tau are only one example of a rich research agenda which a 1–5 µm capability would enable on ATLAST; this agenda will evolve with time in response to the advances JWST will make in many areas of astrophysics, including studies of protoplanetary disks. In Appendix A, we describe the performance of a point design spectrograph designed to work with ATLAST and cover the ~1–5 µm spectral band with resolving power R ~ 200. The sensitivity of this system, reaching S/N > 10 on targets of 15th magnitude at 4 µm in a reasonable integration time, would allow it to obtain high quality spectra of millions of sources catalogued at 3.4 and 4.6 µm by the WISE all sky survey. ATLAST spectra of these sources, ranging from nearby brown dwarfs, young stellar objects such as HL Tau, main sequence and post main sequence stars, stars in the nearest galaxies, and distant galaxies and AGN, could enable a very wide variety of scientific investigations. Within the solar system, spectra extending to 5 μm would allow us to leverage the recent and spectacular New Horizons results on Pluto and Charon in studies of the surfaces of numerous large Trans Neptunian Objects, and also to study the role of sublimating CO and $CO_2$ ices in driving cometary activity in the outer Solar System.

## 6   Summary: An Infrared Capability for ATLAST/LUVOIR

To recapitulate, we have demonstrated that a ~10 m diameter telescope operating in space at room temperature would be able to make unique and important scientific measurements in the infrared out to at least 5 μm. By using the technique of transit spectroscopy, where for bright stars the noise is set by the stellar photons rather than the thermal background, such a telescope could outperform smaller but lower temperature telescopes for studies which could prove critical for the study of exoplanet atmospheres. It would be capable of obtaining spectra of millions of infrared sources from the WISE catalog and provide new information about the formation of solar systems which would link directly to the exoplanet observations. We also have found that, for an ambient temperature telescope, direct imaging and spectrophotometry behind a starlight suppression system provides less capability for exoplanet studies in the 2–5 μm wavelength range than does transit spectroscopy, although the direct imaging approach becomes more powerful with even relatively modest cooling of the telescope. A wider range of parameter space, including extension of the long wavelength limit for the ambient temperature system to at least 8 μm, as well as modest cooling of the telescope mirror to at least below 200 K – and perhaps to ~150 K to achieve zodiacal emission limited performance out to ~4 μm, ought to be considered in more complete studies of both the LUVOIR and the HabEx systems.

To complete this work, we present in Appendix A a conceptual design for a small prism spectrometer which provides R = 200 point source spectroscopy over the 1–5 μm band. R = 200 is appropriate resolution for the study of exoplanet atmospheres and related science themes described above. This point-design is intended as a proof of concept to demonstrate that an instrument which exploits the scientific potential of a warm telescope for infrared work can be installed in the ATLAST focal plane without its cooling requirements compromising the performance of the other instruments which are devoted to shorter wavelength observations. See Werner, Wang, and Macenka[20] for additional information about this spectrometer design. Of course other implementations might meet our scientific objectives; possibilities would include a cryogenically or mechanically-cooled instrument in the main focal plane, similar to NICMOS on HST, or an externally mounted instrument fed by mirrors rather than fibers.

We feel that the scientific argument presented earlier, together with the simple implementation put forward in the Appendix, provide a compelling rationale for including infrared spectroscopy out to at least 5 μm as part of the baseline measurement suite for ATLAST/LUVOIR.

**Appendix A: A Fiber-Fed Infrared Spectrometer for ATLAST/LUVOIR**

*A.1 Introduction*

We describe a fiber-fed IR spectrometer for ATLAST. We have used a radiometry model to estimate the SNR at the final detector plane by calculating the target star signal and various background noises from telescope, fiber, spectrometer, and FPA. The spectral resolving power of 200 is chosen to permit disambiguation of molecular features in exoplanet atmospheres and also to permit study of the profiles of solid state absorption features in galactic protostellar and protoplanetary disks. This report describes the spectrograph design and radiometry model and presents some analysis results. We emphasize that other spectrograph designs and means of feeding the spectrograph could provide equivalent performance, and that coupling the starlight into and out of the fiber could stress the ATLAST pointing system and perhaps other design parameters. An alternative approach using mirrors instead of fibers to feed the spectrograph ought certainly to be considered, for example. This particular approach was chosen because, on the surface at least, it appears to minimize the system impact of the thermal infrared capability.

*A.2 System Configuration*

Figure A-1 is a concept sketch of the system under the analysis. A pickup mirror will be inserted into the ATLAST telescope focal plane (9.2 m aperture, f/12.5) to relay the star signal into a prism-based spectrometer system (f/12.5), featuring a MWIR FPA (5 μm -cutoff H2RG from Teledyne). The reflective prism is made of calcium fluoride. A 10 m IR fiber (Fluoride Fiber ZBLAN SM from Le Verre Fluore) is used to carry the light from the telescope to the spectrometer. To get optimal fiber coupling, 2 relay mirrors with power (f/12.5 to f/2.5) are utilized. To minimize the spectrometer background noise a cold shield box covers the full spectrometer system, and the second relay surface illuminates the spectrometer slit, which is mounted on the side of the otherwise sealed cold shield box.

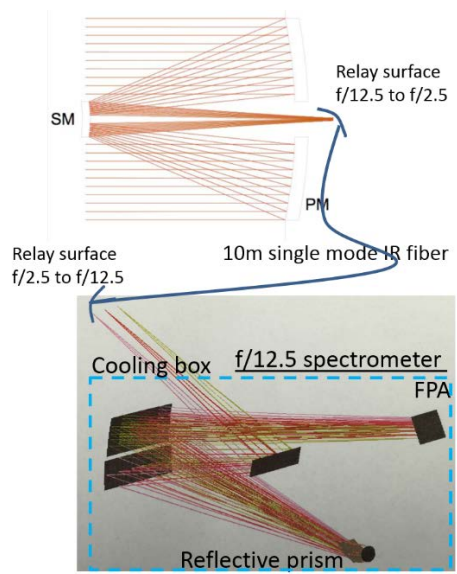

**Fig. A-1** Fiber-fed, prism-based IR spectrometer configuration.

The detailed mechanical packaging of the system has not yet been done, but the optical layout suggests a dimension no greater than 10×10×10 cm—the size of a cubesat—for the entire instrument, including optics, detectors, and electronics. The small volume assures that the instrument can be readily cooled radiatively, as will be required if it is to achieve its expected performance. The concept is that the instrument would be mounted on the telescope backup structure with a clear view to cold space to facilitate the radiative cooling. (We presume that the region of the exterior of the backup structure where the instrument is mounted has itself cooled radiatively, although perhaps not to as low a temperature as we envision for the instrument. If instead it is maintained at or near 300 K, careful engineering and attention to details such as outgassing will be required for the radiatively cooled instrument to achieve satisfactory performance.) Three entrance apertures arranged in a line and some ~15 arcsec apart in the main ATLAST focal plane would couple optically to fibers which would carry the signal to the instrument, where a second lens system would match the light emergent from the fibers to the entrance aperture of the spectrograph. If the star were placed in the central aperture, the spectra obtained through the two lateral apertures would determine the instrumental and telescope background, which could then be subtracted from the spectrum of the central aperture to determine the true stellar spectrum. Details of the radiometry model and some typical results are presented below.

*A.3 Modeling Parameters*

Table A-1 lists the radiometry model parameters used in the analysis.

**Table A-1** Radiometry modeling parameters.

| Item | Value |
|---|---|
| Spectral band | 4 μm center, 0.02 μm bandwidth, spectral resolution of 200 (Fig. A-2). |
| Stellar flux | Assume that a 5th magnitude star has a flux of 2 Jy at 4 μm. Adopt a stellar temperature of 5500 K for wavelength dependent calculations. |
| Telescope | 9.2 m diameter, 15% obscuration, f/# = 12.5, 2 mirrors with each at 0.97 reflectivity and total emissivity of 0.06, T = 293 K. Same parameters except for diameter and temperature adopted for 12 m case. |
| Relay surface[1] | f/12.5 to f/2.5 conversion/relay surface, reflectivity 0.97 and emissivity of 0.03, T = 293 K. |
| Fiber | 10 m long SM 6.5 μm core fiber connects TMA and spectrometer, Mode-filled-diameter (MFD) at 4 μm = 19.7 μm, NA = 0.2, 0.5 dB loss, fiber coupling efficiency 95%, emissivity 0.15, T = 293 K. |
| Relay surface[2] | f/2.5 to f/12.5 conversion/relay surface, reflectivity 0.97 and emissivity of 0.03, T = 293 K. |
| Spectrometer | f/12.5 system, 3 mirrors with each at 0.97 reflectivity, 1 reflecting calcium fluoride prism with 0.9 reflectivity, whole spectrometer included in a cold shield box (blackbody with emissivity of 1), T = 150 K. |
| FPA | 18 μm pitch, 8e4 full well, QE 0.7, 14 bits ADC, 20e-readnoise, T = 80 K, dark current 5 e/s, each resolution element is spread over ~7 × 7 pixels. We note a referee's comment that a lower detector temperature for the % μm HgCdTe focal plane may be required if the unit cell is to function properly as an integrator. |

We present results for two configurations. The baseline is a 9.2 m telescope at 293 K, while the stretch goal is a 12-m telescope at 273 K.

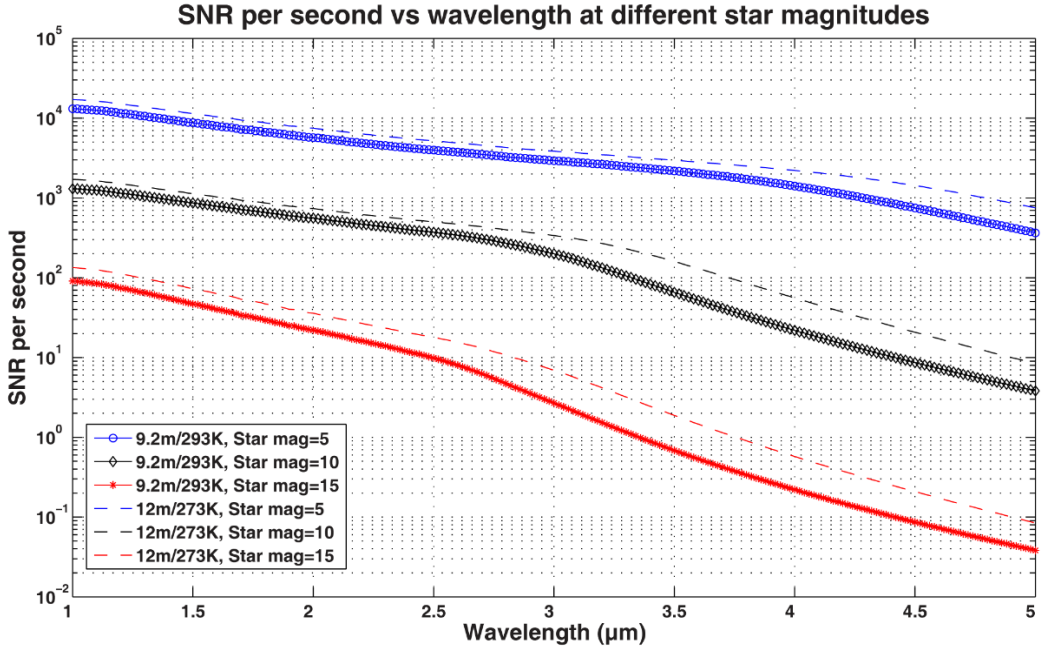

**Fig. A-2** SNR vs star magnitude for two telescope settings.

Figure A-2 shows the S/N, summed over all the pixels in each R = 200 spectral resolution element, as a function of wavelength for several values of telescope aperture and temperature, and for stars of various brightness. A stellar temperature of 5500 K and a blackbody spectrum are assumed, and the SED is fit through the cited magnitude at 4 μm.

**Acknowledgments**

We thank Harley Thronson, Marc Postman, and the other members of the ATLAST study team for their encouragement and support. Aki Roberge provided useful advice and a detailed review of the manuscript. We appreciate the helpful comments of two anonymous referees. We thank Mary Young for editorial assistance. The design of the spectrometer described in Appendix A was supported by a grant from the JPL Research and Technology Development fund. Portions of the work described in this paper were carried out at the Jet Propulsion Laboratory, California Institute

 

**Biographies**

**Dr. Shawn Domagal-Goldman** is an astrobiologist at NASA's Goddard Space flight Center. After receiving degrees in Physics (BS) and Geology (MS) at the University of Rochester, he earned a PhD in Astrobiology and Geosciences at The Pennsylvania State University. Through his career, he has worked on various biosignatures and their false positives, from iron isotopes in the rock record of Mars and Earth to remotely detectable gases on exoplanets.

**Dr. Joel Green** is the Project Scientist in the Office of Public Outreach at Space Telescope Science Institute, home of the science operations of the Hubble and upcoming James Webb Space Telescopes. He is a specialist in the study of young stellar objects and protostars via multi-wavelength spectroscopy, with a particular focus on variability and accretion processes.

**Dr. Avi Mandell** is a civil servant Research Scientist at the Goddard Space Flight Center. He specializes in low- and high-resolution IR spectroscopy of extrasolar planet atmospheres and circumstellar disks, N-body modeling of terrestrial planet formation, and mission development for exoplanet science.

**Dr. Steven Macenka** has contributed in significant ways to establishing and maintaining JPL's world-class expertise and leadership in interferometry and advanced optics. He played a key role in the design and the development and delivery of the Wide-Field Planetary Camera-2 which has resulted in numerous scientific observations and discoveries on the Hubble Space Telescope. Later, he was instrumental in the development, evaluation and verification of the telescope system for the Spitzer Space Telescope.

**Dr. Mark Swain**, who received his Ph.D. in Physics and Astronomy from the University of Rochester in 1996, is a Principal Scientist and Supervisor for the Exoplanet Discovery and Science Group at the Jet Propulsion Laboratory. He is one of the pioneers in detecting molecules in exoplanet atmospheres using near infrared spectroscopy and has extensive experience in infrared instrumentation. His current research interests include comparative exoplanetology and the characterization of potentially habitable worlds.

**Dr. Michael Werner** is a Senior Research Scientist at the Jet Propulsion Laboratory, California Institute of Technology, and Project Scientist for the Spitzer Space Telescope. From 2004 to 2014 he served as Chief Scientist for Astronomy and Physics at JPL. He received NASA's Distinguished Public Service Medal in 2010 and was the George Darwin Lecturer for the Royal Astronomical Society in 2006.

# Tables

**Table 1** Limiting Magnitude for Determination of Molecular Abundances with an Overall Uncertainty of Less Than One Order of Magnitude for a 2.6 $R_{Earth}$ Super Earth/sub-Neptune (T ~ 600 K) orbiting a K star.

|  | Molecule | 0.4–2.5 μm | 0.4–5.0 μm |
|---|---|---|---|
| Mean mol wt 2.3 | $H_2O$ | $H_{mag} < 10.3$ | $H_{mag} < 11.3$ |
|  | $CH_4$ | $H_{mag} < 10.3$ | $H_{mag} < 12.3$ |
| Mean mol wt 28 | $H_2O$ | $H_{mag} < 6.3$ | $H_{mag} < 8.3$ |
|  | $CH_4$ | $H_{mag} < 6.3$ | $H_{mag} < 8.3$ |

**Table A-1** Radiometry modeling parameters.

| Item | Value |
|---|---|
| Spectral band | 4 μm center, 0.02 μm bandwidth, spectral resolution of 200 (Fig. A-2). |
| Stellar flux | Assume that a 5th magnitude star has a flux of 2 Jy at 4 μm. Adopt a stellar temperature of 5500 K for wavelength dependent calculations. |
| Telescope | 9.2 m diameter, 15% obscuration, f/# = 12.5, 2 mirrors with each at 0.97 reflectivity and total emissivity of 0.06, T = 293 K. Same parameters except for diameter and temperature adopted for 12 m case. |
| Relay surface[1] | f/12.5 to f/2.5 conversion/relay surface, reflectivity 0.97 and emissivity of 0.03, T = 293 K. |
| Fiber | 10 m long SM 6.5 μm core fiber connects TMA and spectrometer, Mode-filled-diameter (MFD) at 4 μm = 19.7 μm, NA = 0.2, 0.5 dB loss, fiber coupling efficiency 95%, emissivity 0.15, T = 293 K. |
| Relay surface[2] | f/2.5 to f/12.5 conversion/relay surface, reflectivity 0.97 and emissivity of 0.03, T = 293 K. |
| Spectrometer | f/12.5 system, 3 mirrors with each at 0.97 reflectivity, 1 prism with 0.9 reflectivity, whole spectrometer included in a cold shield box (blackbody with emissivity of 1), T = 150 K. |
| FPA | 18 μm pitch, 8e4 full well, QE 0.7, 14 bits ADC, 20e-readnoise, T = 80 K, dark current 5 e/s, each resolution element is spread over ~7 × 7 pixels. We note a referee's comment that a lower detector temperature for the % μm HgCdTe focal plane may be required if the unit cell is to function properly as an integrator. |

**List of Figure Captions**

**Fig. 1** The time taken (or inverse sensitivity) for three telescope configurations to detect with 5:1 signal-to-noise a small (10 ppm at R = 200) and narrow feature at 4 μm due to a transiting exoplanet, as a function of the stellar magnitude at that wavelength. The configurations are baseline ATLAST, the larger and somewhat cooler "stretch goal" ATLAST, and JWST. ATLAST is more sensitive than JWST for bright host stars, with the 12 m ATLAST telescope taking ~4× less time to make the same observation, as is to be expected in the stellar photon limited case. For the ATLAST configurations, a total photon throughput of 0.35, a (single mode) background with emissivity of 0.2 is assumed; these parameters are commensurate with those of the instrument described in Appendix A. For JWST, the actual expected performance of the NIRCAM grism[8] is adopted.

**Fig. 2** Signal/noise for each resolution element on the transit in one hour for planetary transits at 4 μm. A 9.2 m 293 K telescope and the R = 200, fiber-fed spectrograph described in Appendix A are used for the measurement.

**Fig. 3** An example of abundance estimation from transit spectroscopy. The figure shows a simulated transmission spectrum of a Super Earth/sub-Neptune with high mean molecular weight (mmw = 28) atmosphere orbiting an $H_{mag}$ = 8.3 K dwarf host star. Six transits are coadded—leading to a total integration time of 24 hours—to produce this spectrum with a 9.2 m ATLAST and nominal instrumental parameters. The forward model and retrieval used in this analysis are part of the CHIMERA package developed by Mike Line[16]. The plotted spectrum was obtained with a resolving power of 1000; if the resolution were degraded to the R = 200 level obtainable with the spectrometer described elsewhere in this paper, the abundance determinations for the broad molecular bands would not be materially affected. As summarized in Table 1, spectra of this quality extended to 5 μm and analyzed with CHIMERA can constrain the abundances of molecules like $CH_4$ and $H_2O$ in exoplanet atmospheres with a total uncertainty less than an order of magnitude.

**Fig. 4** A 2–5 μm R ~ 20 model transmission spectrum of a 1 $R_{Earth}$ habitable zone planet with an Earth-like oxidizing atmosphere transiting an M4V star ($R_{star}$ = 0.32 $R_s$, $T_{star}$ = 3100 K) which is 15 pc from Earth. The 1-sigma errors per wavelength bin, obtained with the baseline ATLAST design (D = 9.2 m, T = 273 K), are plotted along with the model. The total integration time is 120 hours in transit, corresponding to 60 transits of about 2 hours each. Features of $H_2O$, $CH_4$, and $CO_2$ are detected. $CH_4$ is blended with $H_2O$ at ~3.3 μm and determining its independent presence and total column would require full atmospheric retrieval; in the Earth's atmosphere, $CH_4$ and $H_2O$ contribute roughly equally to the 3.3 μm feature. A total throughput of 0.35 and an emissivity of 0.2 is assumed here. The y-axis is labelled in parts per million relative to the total signal produced by the star. The overall depth of the transit for this case would be about 800 ppm; the features shown in the spectrum would be manifest as changes in the transit depth with wavelength.

**Fig. 5** A hypothetical water (80%) and $CO_2$ (20%) (mean molecular weight of 23) sub-Neptune planet transiting the same M4V star (15 pc). This planet has a radius of 2.7 $R_{Earth}$ and a mass of 6.5 $M_{Earth}$. The atmospheric temperature is ~500 K, so it is a bit closer to the star than is the Earth analogue shown in Fig. 5. Error bars indicate the 1-sigma precision that could be obtained along the spectrum per R = 100 spectral bin in 6 hours, which would require observations of a handful of transits.

**Fig. 6** Number of HZs potentially containing Earth analogs which could be characterized at 2.3 μm (orange) and 3.4 μm (red) as a function of telescope temperature, assuming an upper limit of 100 days of exposure time on each system. Cooling to ~275 K provides access to the Alpha Cen binary system. Cooler temperatures are required to reduce the exposure times of other systems. A 12-m aperture telescope is assumed.

**Fig. 7** HL Tau as seen by ALMA (https://public.nrao.edu/news/pressreleases/planet-formation-alma), compared with approximate JWST and HDST beamsizes at 3 μm.

**Fig. A-1** Fiber-fed, prism-based IR spectrometer configuration.

**Fig. A-2** SNR vs star magnitude for two telescope settings.